\documentclass[12pt]{article}
\usepackage{epsf}
\title{ {\bf
Annihilation cross sections and interaction couplings of the dark
matter candidates in the warped and flat extra dimensions}}
\author{\vspace{1cm}\\
        {\bf E. O. Iltan}
        \thanks{E-mail address:
        eiltan@newton.physics.metu.edu.tr}
\\
Physics Department, Middle East Technical University \\
        Ankara, Turkey \\
        }

\date{}

\begin{document}
\setlength{\baselineskip}{24pt}
\maketitle
\setlength{\baselineskip}{7mm}
\begin{abstract}
We consider a scenario with an additional scalar standard model
singlet $\phi_S$, living in a single extra dimension of the RS1
background. The zero mode of this scalar which is localized in the
extra dimension is a dark matter candidate and the annihilation
cross section is strongly sensitive to its localization parameter.
As a second scenario, we assume that the standard model Higgs
field is accessible to the fifth flat extra dimension. At first we
take the additional standard model singlet scalar field as
accessible to the sixth extra dimension and its zero mode is a
possible dark matter candidate. Second, we consider that the new
standard model singlet, the dark matter candidate, lives in four
dimensions. In both choices the KK modes of the standard model
Higgs field play an observable role for the large values of the
compactification radius $R$ and the effective coupling
$\lambda_{S}$ is of the order of $10^{-2}-10^{-1}$ ($10^{-6}$) far
from (near to) the resonant annihilation.
\end{abstract}
\thispagestyle{empty}
\newpage
\setcounter{page}{1}
The missing matter which is required holds almost $23\%$
\cite{JungmanG, JungmanG2, KomatsuE} of present Universe and it is
called dark matter (DM) since it is not detectable by the
radiation emitted. The evidence of the existence of DM comes from
numerous observations: the galactic rotation curves
\cite{BorrielloA}, galaxies orbital velocities \cite{ZwickyF}, the
cosmic microwave background anisotrophy \cite{WAMP2}, observations
of type Ia supernova \cite{KomatsuE}. However the nature of  DM is
still a mystery. The DM problem can not be solved in the framework
of the standard model (SM) and it is inevitable to search new
physics beyond in order to provide a dark matter candidate. In the
literature, there are many studies which are based on the models
beyond the SM in order to understand the nature of DM; DM in the
framework of the supersymmetric models \cite{MAsano}, the
universal and non universal extra dimension (UED and NUED) models
\cite{Cheng1}-\cite{Eiltan1}, the split UED models \cite{ServantG,
Park, Nojiri}, the Private Higgs model \cite{Jackson}, the Inert
doublet model \cite{MaE}-\cite{Calmet}, the Little Higgs model
\cite{Bai}, the Heavy Higgs model \cite{Majumdar}. The common idea
is that a large amount of DM is in the class of nonrelativistic
cold DM and the Weakly Interacting Massive Particles (WIMPs)
belong to this class. WIMPs, having masses in the range $10$ GeV-
a few TeV, involve in the weak and gravitational interactions and
they are stable in the sense that they do not decay in to SM
particles and they play a crucial role in the structure formation
of Universe. On the other hand they disappear by pair annihilation
(see for example \cite{DEramoF, WanLeiGuo} for further
discussion). Notice that the stability of WIMPs are ensured by an
appropriate discrete symmetry, in various models (for details see
for example \cite{ChuanRenChen} and references therein).

From the experimental point of view there are two possibilities to
detect the DM candidate WIMP: The direct detection of DM, the
search for the scattering of DM particles off atomic nuclei within
a detector, and the indirect one, the search for products of WIMP
annihilations. An upper limit of the order of $10^{-7}-10^{-6}\,
pb $ \cite{Akerib} for the WIMP-nucleon cross section has been
obtained in the direct detection experiments. On the other hand
since the current relic density could be explained by thermal
freeze-out of their pair annihilation the present DM abundance by
the WMAP collaboration \cite{WAMP} leads to the bounds for the
annihilation cross section.

In the present work we study the annihilation cross section and
the related coupling of the DM candidates in the warped and flat
extra dimensions. At first, we consider that all SM particles live
on the 4 dimensional brane and there exists an additional scalar
SM singlet $\phi_S$ which is accessible to a single extra
dimension in the RS1 background and its zero mode is a possible DM
candidate. As a second scenario we choose the flat extra
dimension(s) where the SM Higgs doublet, necessarily the gauge
fields, are accessible to a single extra dimension (the fifth one)
with two possibilities: the additional SM model singlet scalar
field lives in the sixth extra dimension and its zero mode is a
possible DM candidate; the new SM singlet, the DM candidate, lives
in four dimensions.
%
\\ \\
{\Large \textbf{DM as the zero mode of SM singlet $\phi_S$ which
is accessible to a single extra dimension in the RS1 background}}
\\ \\
The RS1 background\footnote{Here, the extra dimension, having two
boundaries, the hidden (Planck) brane and the visible (TeV) brane
with opposite and equal tensions, is compactified onto $S^1$
orbifold with the compactification radius $R$. In this case the
low energy effective theory has flat 4D spacetime, even if the 5D
cosmological constant is non vanishing. The gravity, having an
extension into the bulk with varying strength, is taken to be
localized on the hidden brane.}  \cite{Rs1, Rs2} is based on the
curved extra dimension with the metric
\begin{eqnarray}
ds^2=e^{-2\,\sigma}\,\eta_{\mu\nu}\,dx^\mu\,dx^\nu-dy^2\, ,
\label{metric1}
\end{eqnarray}
where $\sigma=k\,|y|$, $k$ is the bulk curvature constant, the
exponential $e^{-\sigma}$, with $y=R\,|\theta|$,  is the warp
factor. Now we consider that an additional SM singlet $\phi_S$ is
accessible to the extra dimension. The compactification of the
extra dimension  onto $S^1$ orbifold with radius $R$ results in
the appearance of KK modes as
\begin{eqnarray}
\phi_S(x,y)=\sum_{n=0}^\infty\, \phi_S^{(n)}(x)\, f_n(y) \, ,
\label{phiKK}
\end{eqnarray}
where the non-vanishing zero mode exists if the  fine tuning of
the parameters (see \cite{Goldberger, Kogan} and \cite{Eiltan2}
for details)\footnote{There is another possibility of fine tuning
of the parameters  $b$ and $a$ for the non-vanishing zero mode,
namely $b=2+\sqrt{4+a}$. However we ignore this choice since it is
not appropriate for our case since it leads to an effective
coupling which is not valid for the perturbative calculation.} $b$
and $a$
\begin{eqnarray}
b=2-\sqrt{4+a}
 \, ,\label{finetune}
\end{eqnarray}
is reached, and it reads
\begin{eqnarray}
f_0(y)=\frac{e^{b\,k\,y}}{\sqrt{\frac{e^{2\,(b-1)\,
k\,\pi\,R}-1}{(b-1)\,k}}}
 \,\, . \label{f0}
\end{eqnarray}
If one respects the existence of the ad-hoc discrete $Z_2$
symmetry $\phi_S\rightarrow -\phi_S$ and considers that $\phi_S$
has no vacuum expectation value, the stability of the zero mode
scalar $\phi^{(0)}_S$ is guaranteed\footnote{The scalar field
$\phi_S$ has no SM decay products.} and it can be taken as a DM
candidate which disappears by pair annihilation with the help of
the exchange particle. The possible interaction which drives the
pair annihilation is
\begin{eqnarray}
{\cal{S}}_{Int}= \int d^5x \sqrt{g} \,\Bigg(
\lambda_{5\,S}\,(\phi_S^{(0)})^2\,(\Phi_1^\dagger\,
\Phi_1)\,\Bigg)\, \delta(y-\pi R) \,\,\, ,\label{VintRS}
\end{eqnarray}
where $\Phi_1$ is the SM Higgs field
\begin{eqnarray}
\Phi_{1}=\frac{1}{\sqrt{2}}\left[\left(\begin{array}{c c}
0\\v+H^{0}\end{array}\right)\; + \left(\begin{array}{c c} \sqrt{2}
\chi^{+}\\ i \chi^{0}\end{array}\right) \right]\, ,
\label{HiggsField}
\end{eqnarray}
with the vacuum expectation value
\begin{eqnarray}
<\Phi_{1}>=\frac{1}{\sqrt{2}}\left(\begin{array}{c c}
0\\v\end{array}\right) \, . \label{VEV}
\end{eqnarray}
Here the SM Higgs boson $H^0$ is the exchange particle and the
pair annihilation occurs after the electroweak symmetry breaking.
In this part of the work we will study the effects of the zero
mode scalar localization parameter $a$ and the curvature $k$ on
the annihilation cross section.
%
\\ \\
{\Large \textbf{DM as the SM singlet (or the zero mode of the SM
singlet living in the sixth flat extra dimension) and the Higgs
field living in the fifth flat extra dimension}}
\\ \\
At first, we assume that the SM Higgs doublet and the additional
SM model singlet scalar field are accessible to fifth and sixth
extra dimension respectively, however the SM fields, except gauge
fields, live in four dimensions. The compactification of the extra
dimensions on $S_1\times S_1$ with radii $R$ results in the
expansion of the SM Higgs doublet $\Phi_1$ (see
eq.(\ref{HiggsField})) and the new SM singlet $\phi_S$ into their
KK modes as
\begin{eqnarray}
\Phi_1(x,y) & = & {1 \over {\sqrt{2 \pi R}}} \left\{
\Phi_1^{(0)}(x) + \sqrt{2} \sum_{n=1}^{\infty} \Phi_1^{(n)}(x)
\cos(ny/R)\right\} \,,
\nonumber \\
\phi_S (x,y)& = & {1 \over {\sqrt{2 \pi R}}} \left\{
\phi_S^{(0)}(x) + \sqrt{2} \sum_{n=1}^{\infty} \phi_S^{(n)}(x)
\cos(nz/R)\right\} \,, \label{SecHiggsField}
\end{eqnarray}
where  $y$ and $z$ are the coordinates of the fifth and sixth
extra dimensions. Now, we consider the interaction of the
additional scalar singlet with the SM Higgs doublet as
\begin{eqnarray}
{\cal{L}}_{Int}= \Bigg(
\lambda_{6\,S}\,\phi_S^2\,(\Phi_1^\dagger\, \Phi_1)\,\Bigg)_{y=0,
z=0}\, .\label{Vint}
\end{eqnarray}
After the electroweak symmetry one gets the interaction term
\begin{eqnarray}
{\cal{L^\prime}}_{Int}=\frac{\lambda_{6\,S}\,v}{(2\,\pi\,R)^2}\,
(\phi_S^{(0)})^2\,\Bigg(
H^{0\,(0)}+\sqrt{2}\,\sum_{n=1}\,H^{0\,(n)} \Bigg)\,
,\label{Vintp}
\end{eqnarray}
which is responsible for the the annihilation process of
$\phi_S^{(0)}$ which we consider as a DM candidate. Here the zero
mode and KK mode Higgs fields are intermediate particles which
carry the annihilation process. Notice that the stability of the
DM candidate under consideration is ensured by respecting that the
SM singlet scalar $\phi_S$, having no vacuum expectation value,
obeys the discrete $Z_2$ symmetry $\phi_S\rightarrow -\phi_S$.

Second, we consider that the SM Higgs doublet is accessible to
fifth extra dimension, however, the additional SM model singlet
scalar field, the DM candidate, lives in four dimensions. This is
the case that the interaction of the additional scalar singlet
with the SM Higgs doublet reads
\begin{eqnarray}
{\cal{L}}_{Int}= \Bigg(
\lambda_{5\,S}\,\phi_S^2\,(\Phi_1^\dagger\,
\Phi_1)\,\Bigg)_{y=0}\, ,\label{Vint5}
\end{eqnarray}
and the interaction term, which is responsible for the
annihilation process, becomes
\begin{eqnarray}
{\cal{L^\prime}}_{Int}=\frac{\lambda_{5\,S}\,v}{2\,\pi\,R}\,
\phi_S^2\,\Bigg( H^{0\,(0)}+\sqrt{2}\,\sum_{n=1}\,H^{0\,(n)}
\Bigg)\, ,\label{Vint5p}
\end{eqnarray}
after the electroweak symmetry breaking. Similar to the previous
case the zero mode and KK mode Higgs fields play the role of
intermediate particles which drive the annihilation process. The
stability of the DM candidate is ensured with the above ad-hoc
$Z_2$ symmetry and with vanishing vacuum expectation value.

Now, we present the total averaging annihilation rate of DM which
is obtained by the annihilation process DM DM$\rightarrow
H^0\,\rightarrow X_{SM}$
\begin{eqnarray}
<\sigma\,v_r>&=&\frac{4\,\lambda_S^2\, v^2}{m_S}\,\frac{1}{
(4\,m_S^2-m_{H^0}^2)^2+m^2_{H^0}\,\Gamma^2_{H^0}}\,
\Gamma(\tilde{h}\rightarrow X_{SM})\, ,  \label{sigmavr}
\end{eqnarray}
where $\Gamma(\tilde{h}\rightarrow
X_{SM})=\sum_i\,\Gamma(\tilde{h}\rightarrow X_{i\,SM})$ with
virtual Higgs $\tilde{h}$ having mass $2\,m_S$ (see \cite{BirdC,
BirdC2}) and $v_r=\frac{2\,p_{CM}}{m_S}$ is the average relative
speed of two zero mode scalars (see for example \cite{HeXG}). Here
the effective coupling $\lambda_S$ model dependent and, for the
case that the DM is the zero mode SM singlet $\phi_S$ which is
accessible to a single extra dimension in the RS1 background, it
reads
\begin{eqnarray}
\lambda_S=\lambda_{5\,S}\,e^{-2\,k\,\pi\,R}\,f^2_0(\pi\,R) \, ,
\label{lambdaSRS}
\end{eqnarray}
where $f_0(y)$ is given in eq.(\ref{f0}). In the case that the SM
Higgs field is accessible to the fifth flat extra dimension and
the DM candidate is the zero mode of the SM singlet, which is
accessible to the sixth one, the total averaging annihilation rate
of DM reads
\begin{eqnarray}
<\sigma\,v_r>&=&\frac{4\,\lambda_S^2\,
v^2}{m_S}\,\Bigg|\,\frac{1}{
(4\,m_S^2-m_{H^{0\,(0)}}^2)+i\,m_{H^{0\,(0)}}\,\Gamma_{H^{0\,(0)}}}\nonumber
\\ &+& \sqrt{2}\,\sum_{n=1}\,\frac{1}{(4\,m_S^2-m_{H^{0\,(n)}}^2)+i\,
m_{H^{0\,(n)}}\,\Gamma_{H^{0\,(n)}}}\,\Bigg |^2\,
\Gamma(\tilde{h}\rightarrow X_{SM})\,,   \label{sigmavr2}
\end{eqnarray}
where $m_{H^{0\,(n)}}^2=m_{H^{0\,(0)}}^2+\frac{n^2}{R^2}$ and
$\lambda_S$ is
\begin{eqnarray}
\lambda_S=\frac{\lambda_{6\,S}}{(2\,\pi\,R)^2} \,
.\label{lambdaS3}
\end{eqnarray}
If the DM candidate is the SM singlet, living in four dimensions,
$\lambda_S$ becomes
\begin{eqnarray}
\lambda_S=\frac{\lambda_{5\,S}}{2\,\pi\,R} \, , \label{lambdaS2}
\end{eqnarray}
and the annihilation cross section is given in
eq.(\ref{sigmavr2}).

For the the annihilation cross section $< \sigma\,v_r>$ we respect
the restriction
\begin{eqnarray}
< \sigma\,v_r>=0.8\pm 0.1 \, pb \,,
\end{eqnarray}
which is constructed in the case that  s-wave annihilation is
dominant (see \cite{KolbEW} for details.). These bounds are coming
from the relic density
\begin{eqnarray}
\Omega\,h^2=\frac{x_f\,10^{-11}\,GeV^{-2}}{< \sigma\,v_r>}
\,,\label{omegahsig}
\end{eqnarray}
where $x_f\sim 25$ \cite{JungmanG2, ServantG, HeXG, Gopalakrishna,
Gopalakrishna2} and, by the WMAP collaboration \cite{WAMP}, the
present DM abundance reads
\begin{eqnarray}
\Omega\,h^2=0.111\pm 0.018 \, . \label{RelDens}
\end{eqnarray}
%
\\ \\
{\Large \textbf{Discussion}}
\\ \\
The present work is devoted to the analysis of the annihilation
cross sections of some DM candidates and the couplings that drive
these cross sections. Here we consider two scenarios. As a first
one, we assume that all SM particles live on the 4 dimensional
brane and there exists an additional scalar SM singlet $\phi_S$
which is accessible to a single extra dimension in the RS1
background. In this case the zero mode of the scalar singlet is a
candidate of DM and it is localized in the extra dimension (see
eq.({\ref{f0}})). The interaction term, which is represented by
the action in eq.(\ref{VintRS}), is responsible for the existence
of the vertex  DM DM $H^0$ which appears after the elecroweak
symmetry breaking. This term drives the annihilation cross section
which should be compatible with the present observed DM relic
density (eq.(\ref{RelDens})) and the strength of the interaction
is regulated by the the effective coupling $\lambda_S$
(eq.(\ref{lambdaSRS})). The free parameters in this scenario are
the Higgs mass $m_{H^0}$, the zero mode scalar mass $m_S$, the
curvature $k$ and the parameter $a$ which plays an essential role
in the localization of DM. In our numerical calculations we take
Higgs mass around $110-120\,GeV$, the DM candidate mass in the
range of $10-80\, GeV$ and we choose two different values for the
curvature $k$, $k=10^7\,GeV$ and $k=10^8\,GeV$. Now we study the
localization parameter $a$ dependence of the annihilation cross
section $< \sigma\,v_r>$  and we estimate the range of $a$ by
respecting the upper and lower bounds of the current experimental
value of the relic abundance, namely $0.7\,pb \leq \,\,<
\sigma\,v_r> \,\,\,\leq 0.9\,pb$.

\begin{figure}[htb]
\vskip -3.0truein \centering \epsfxsize=6.8in
\leavevmode\epsffile{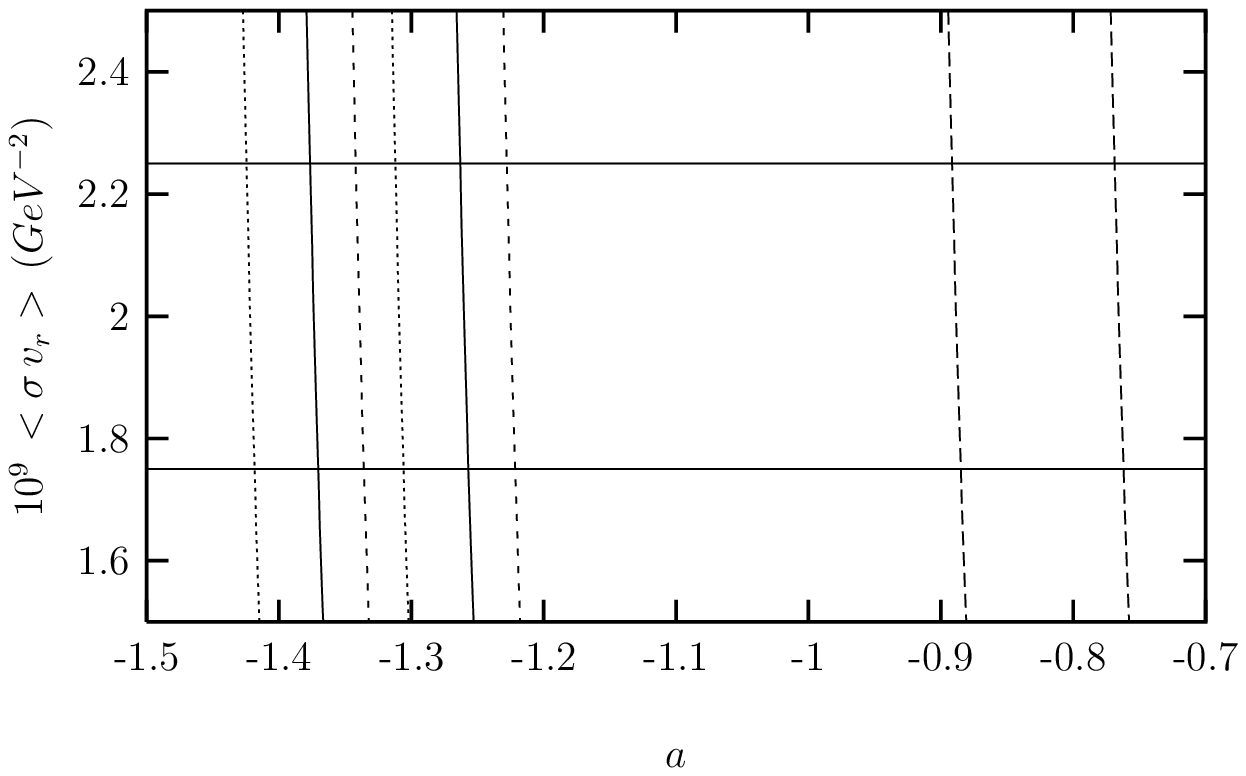} \vskip -3.0truein
\caption[]{ $< \sigma\,v_r>$  as a function of $a$ for
$m_{H^0}=110\,GeV$. Here the left-right solid (long dashed;
dashed; dotted) line represents $< \sigma\,v_r>$ for
$k=10^{17}-10^{18}\,GeV$ $m_S=80\,(55; 50; 10)\,GeV$.}
\label{sigmaversusaRS110x}
\end{figure}

\begin{figure}[htb]
\vskip -3.0truein \centering \epsfxsize=6.8in
\leavevmode\epsffile{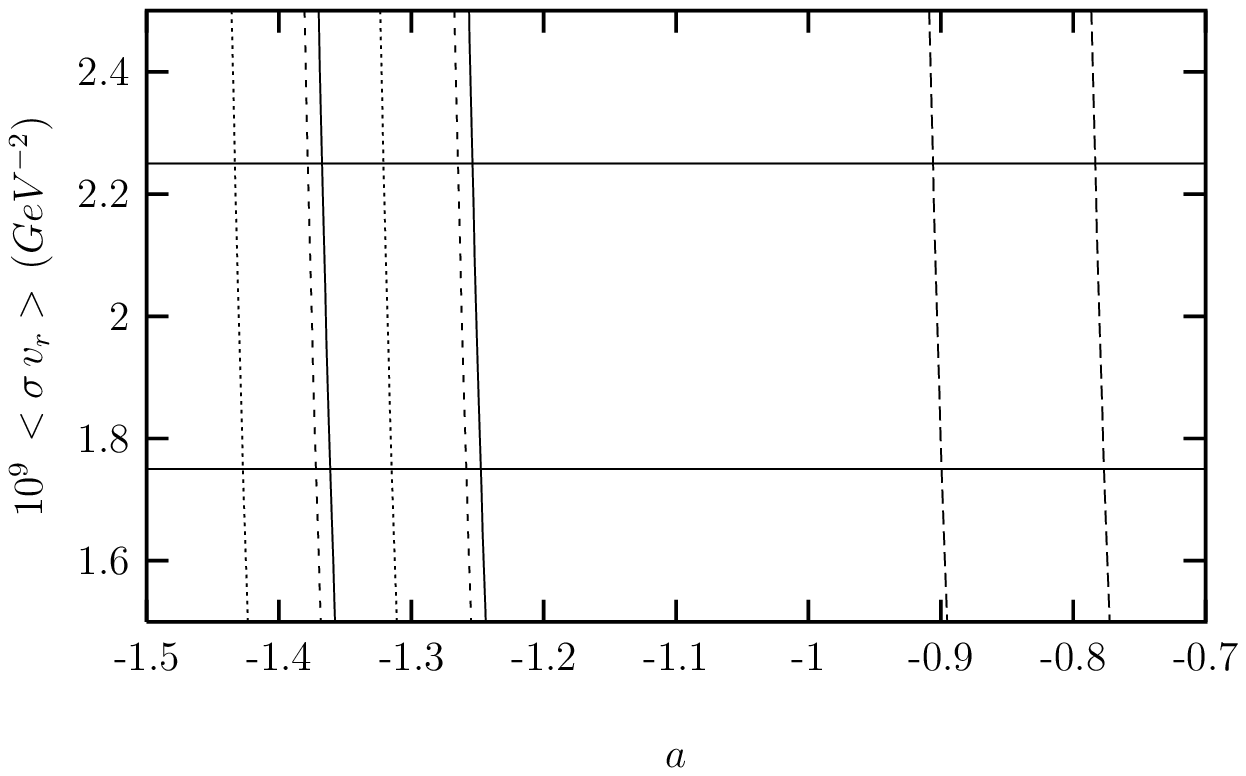} \vskip -3.0truein
\caption[]{The same as Fig\ref{sigmaversusaRS110x} but for
$m_{H^0}=120\,GeV$ and $m_S=80\,(60; 50; 10)\,GeV$.}
\label{sigmaversusaRS120x}
\end{figure}

In Figs.\ref{sigmaversusaRS110x} and \ref{sigmaversusaRS120x} we
plot the localization parameter $a$ dependence of the annihilation
cross section $< \sigma\,v_r>$ for $m_{H^0}=110\,GeV$ and
$m_{H^0}=120\,GeV$. Here the left-right solid (long dashed;
dashed; dotted) line represents $< \sigma\,v_r>$ for
$k=10^{17}-10^{18}\,GeV$ $m_S=80\,(m_R; 50; 10)\,GeV$ where
$m_R=55(60)\,GeV$ for $m_{H^0}=110\,GeV$ ($m_{H^0}=120\,GeV$). We
observe that the annihilation cross section strongly depends on
the parameter $a$. The $0.5\, \%$ variation in $a$ results in that
$< \sigma\,v_r>$ changes between the estimated upper and lower
bounds. If the mass of the scalar is $m_S=55\, GeV$ the resonant
annihilation occurs and $a$ reaches the greatest value so that the
increase in $< \sigma\,v_r>$ is appropriately suppressed in order
to set it in the estimated range. For $m_S=50\, GeV$ $a$ is still
larger compared to ones for $m_S=80\, GeV$ and $m_S=10\, GeV$
since this is the case that the scalar mass is near to the
resonant annihilation mass. For far from resonant annihilation,
heavy scalar mass causes that the ratio in the definition of the
annihilation cross section decreases and the parameter $a$ must
increase to set the cross section in the region which is
restricted by the experimental result (see the curves for
$m_S=80\, GeV$ and $m_S=10\, GeV$). On the other hand the increase
in the compactification radius $R$ results in suppression in the
parameter $a$. For the SM Higgs mass $m_{H^0}=120\,GeV$ the
behavior of the annihilation cross section $< \sigma\,v_r>$ is
similar to the previous case. Here the curve for $m_S=80\, GeV$
lags the one for $m_S=50\, GeV$ since, in this case, the DM scalar
with mass $m_S=50\, GeV$ is relatively far from the resonant
annihilation.

As a second scenario we take the extra dimension(s) flat and, at
first, we assume that the SM Higgs doublet and the additional SM
model singlet scalar field are accessible to fifth and sixth extra
dimension respectively. Here the zero mode of SM singlet is
considered as the DM candidate. This is the case that the
annihilation of the DM occurs with the help of the SM Higgs boson
and its KK modes after the electroweak symmetry breaking (see
eq.(\ref{Vintp})). In this scenario we study the behavior of the
coupling $\lambda_{6S}$ in six dimensions with respect to the
compactification radius $R$, by respecting the current average
value of the annihilation cross section, $< \sigma\,v_r>=0.8\,pb$.
In the numerical calculations we take the compactification radius
$R$ in the range $0.00001\,GeV^{-1} \leq \, R\,
\leq\,0.005\,GeV^{-1}$.

\begin{figure}[htb]
\vskip -3.0truein \centering \epsfxsize=6.8in
\leavevmode\epsffile{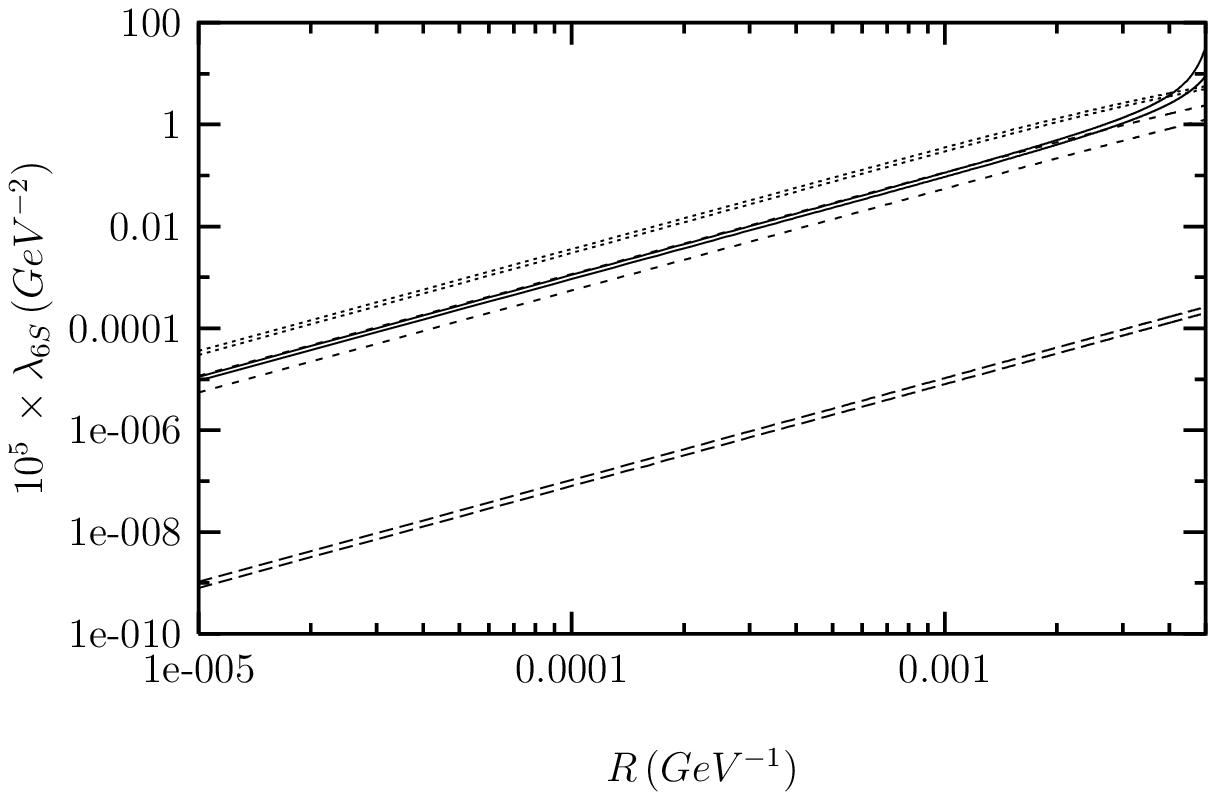} \vskip -3.0truein
\caption[]{ $\lambda_{6S}$ as a function of $R$. Here the
upper-lower solid line represents $\lambda_{6S}$ for
$m_{H^0}=110-120\,GeV$ $m_S=80\,GeV$ and the upper-lower long
dashed (dashed; dotted) line represents $\lambda_{6S}$ for
$m_{H^0}=120-110\,GeV$ $m_S=60-55\,(50; 10)\,GeV$.}
\label{lam6versusRNUED110120}
\end{figure}

Fig.\ref{lam6versusRNUED110120} represents $R$ dependence of the
coupling $\lambda_{6S}$ in six dimension for\footnote{Notice that,
in the following,  we denote the zero mode $H^{0\,(0)}$ as
$H^{0}$.} $m_{H^0}=110\,GeV$ and $m_{H^0}=120\,GeV$. Here the
upper-lower solid line represents $\lambda_{6S}$ for
$m_{H^0}=110-120\,GeV$ $m_S=80\,GeV$ and the upper-lower long
dashed (dashed; dotted) line represents $\lambda_{6S}$ for
$m_{H^0}=120-110\,GeV$ $m_S=60-55\,(50; 10)\,GeV$. $\lambda_{6S}$
lies in the range of $10^{-10}-10^{-6}\,GeV^{-2}$ for the interval
of the compactification radius $10^{-5}-10^{-3}\,GeV^{-1}$ for DM
that is far from the resonant annihilation case. In the case of
resonant annihilation, $\lambda_{6S}$ is suppressed and it is in
the range of $10^{-14}-10^{-10}\,GeV^{-2}$. $\lambda_{6S}$ is
weakly sensitive to the SM Higgs mass for the interval under
consideration, i.e. $110\,GeV \leq m_{H^0}\leq 120\,GeV$. It is
observed that the coupling $\lambda_{6S}$ changes its behavior for
the large values of $R$, $R>0.004\,GeV^{-1}$, especially in the
case that the mass of the DM scalar is far from the resonant
annihilation. This variation comes from the readable effects of
the intermediate SM Higgs KK modes for the compactification radius
in this range. Notice that the Higgs KK mode contribution is
suppressed with the decreasing values of the radius $R$.

\begin{figure}[htb]
\vskip -3.0truein \centering \epsfxsize=6.8in
\leavevmode\epsffile{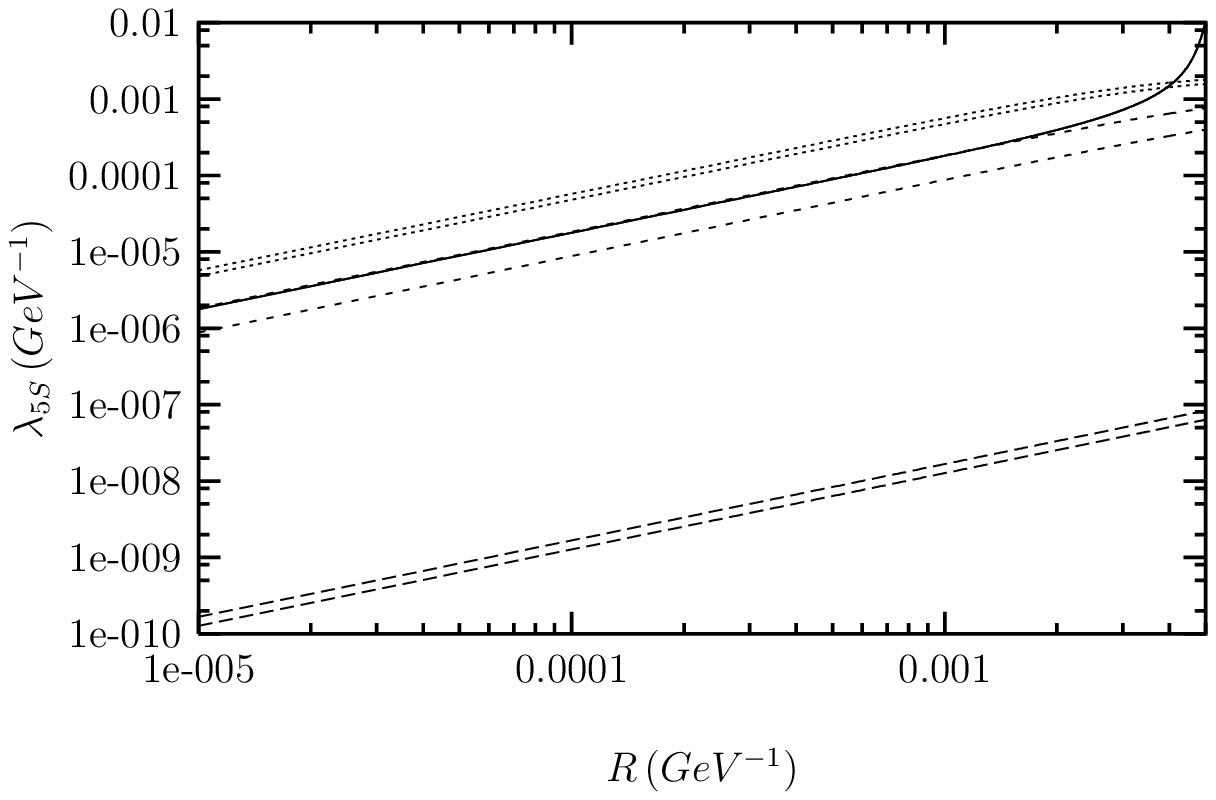} \vskip -3.0truein
\caption[]{The same as Fig\ref{lam6versusRNUED110120} but for
$\lambda_{5S}$.} \label{lam5versusRNUED110120}
\end{figure}

As a second choice, we assume that the SM Higgs doublet is
accessible to fifth extra dimension and the additional SM model
singlet lives in four dimensions. Here we study the behavior of
the coupling $\lambda_{5S}$ with respect to the compactification
radius $R$, by respecting the current average value of the
annihilation cross section, $< \sigma\,v_r>=0.8\,pb$, similar to
the previous case. In Fig.\ref{lam5versusRNUED110120} we show the
$R$ dependence of the coupling $\lambda_{5S}$ for
$m_{H^0}=110\,GeV$ and $m_{H^0}=120\,GeV$. Here the upper-lower
solid line represents $\lambda_{5S}$ for $m_{H^0}=110-120\,GeV$
$m_S=80\,GeV$ and the upper-lower long dashed (dashed; dotted)
line represents $\lambda_{5S}$ for $m_{H^0}=120-110\,GeV$
$m_S=60-55\,(50; 10)\,GeV$. $\lambda_{5S}$ is in the range of
$10^{-6}-10^{-3}\,GeV^{-1}$ for the interval of the radius $R$
$10^{-5}-10^{-3}\,GeV^{-1}$ in the case that the DM mass is far
from the resonant annihilation. If the resonant annihilation
occurs $\lambda_{5S}$ decreases up to the range of
$10^{-10}-10^{-8}\,GeV^{-1}$.  The effects of Higgs KK modes are
observed if the mass of the DM scalar is far from the resonant
annihilation and the radius $R$ is large, $R>0.003\,GeV^{-1}$.

\begin{figure}[htb]
\vskip -3.0truein \centering \epsfxsize=6.8in
\leavevmode\epsffile{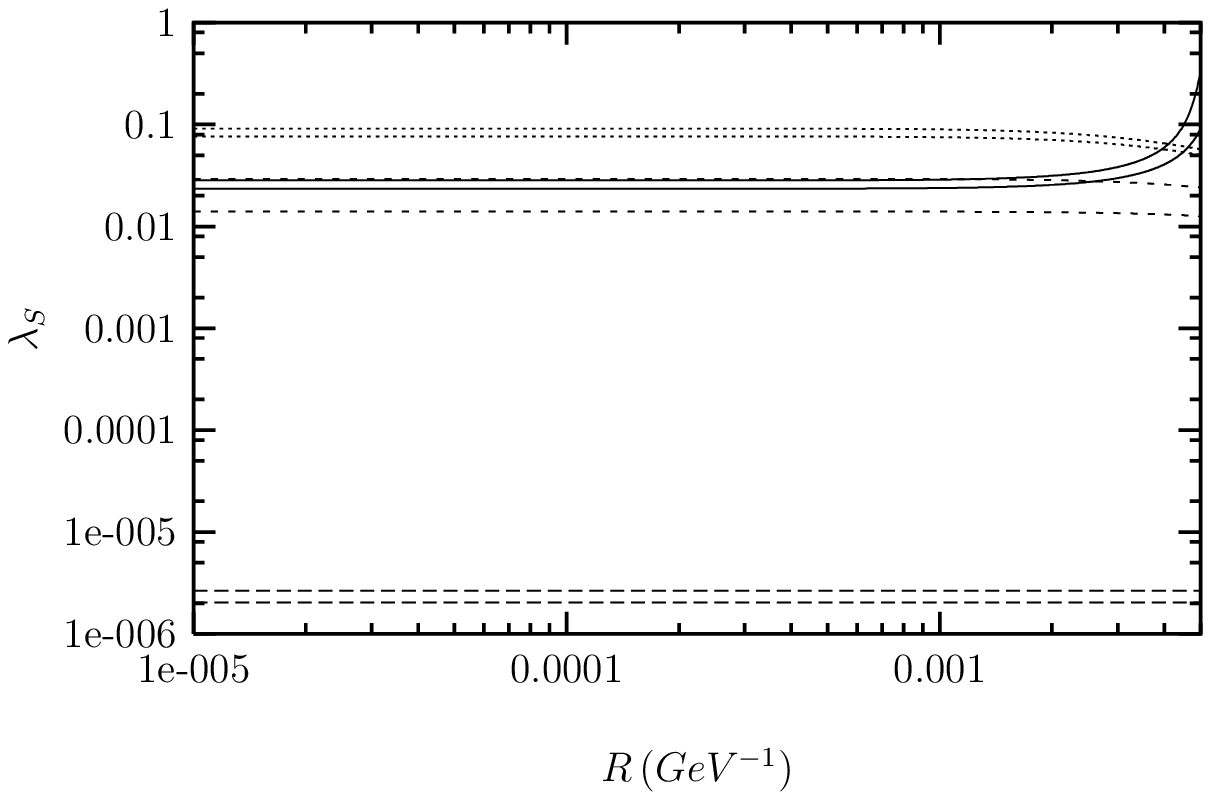} \vskip -3.0truein
\caption[]{The same as Fig\ref{lam6versusRNUED110120} but for
$\lambda_{S}$.} \label{lamEffversusRNUED110120}
\end{figure}

Finally, we plot the the effective coupling $\lambda_{S}$ for both
choices in the second scenario in Fig
\ref{lamEffversusRNUED110120}. Here the upper-lower solid line
represents $\lambda_{S}$ for $m_{H^0}=110-120\,GeV$ $m_S=80\,GeV$
and the upper-lower long dashed (dashed; dotted) line represents
$\lambda_{S}$ for $m_{H^0}=120-110\,GeV$ $m_S=60-55\,(50;
10)\,GeV$. $\lambda_{S}$ is at the order of magnitude of
$10^{-2}-10^{-1}$ ($10^{-6}$) far from (near to) the resonant
annihilation. The effects of intermediate KK modes appear for
$R>0.002\, GeV^{-1}$ and these effects are negligible for the
resonant annihilation case.

At this stage we would like to present our results
\begin{itemize}
\item In the first scenario, the annihilation cross section is
strongly sensitive to the localization parameter $a$ and $a$
reaches its greatest value in the resonant annihilation case. The
increase in curvature $k$ (or the decrease in the compactification
radius $R$) forces $a$ to be suppressed.
\item In the second scenario, we choose the extra dimension(s)
flat and we assume that the SM Higgs doublet is accessible to
fifth dimension. Here we consider two different possibilities. In
the first we take the additional SM model singlet scalar field
which is accessible to the sixth extra dimension and its zero mode
is a possible DM candidate. In the second we consider that the new
SM singlet, the DM candidate, lives in four dimensions. In both
possibilities the KK modes of SM Higgs field play an observable
role for large values of the compactification radius $R$,
$R>0.003\,GeV^{-1}$. On the other hand the dimensionfull couplings
($\lambda_{6S}$ for the first choice and $\lambda_{5S}$ for the
second choice) are weak and the effective coupling $\lambda_{S}$,
which is the same for both choices, is of the order of
$10^{-2}-10^{-1}$ ($10^{-6}$) far from (near to) the resonant
annihilation.
\end{itemize}

The forthcoming more accurate experimental measurements and the
possible observation of the SM Higgs boson at LHC will shed light
on the nature of the DM and its annihilation mechanism.
%
\newpage
\newpage
\end{document}